\documentclass{aa}  

\usepackage{graphicx}
\usepackage{txfonts}
\usepackage{multirow}
\usepackage{array}
\usepackage{xcolor}
\usepackage{soul}

\begin{document}

   \title{Europium, we have a problem} 
   
   \subtitle{Modelling r-process enrichment across Local Group galaxies}

   \author{M. Palla\inst{1,2}
          \and
          M. Molero\inst{3} 
          \and 
          D. Romano\inst{2}
          \and
          A. Mucciarelli\inst{1,2}
          }

   \institute{
            Dipartimento di Fisica e Astronomia “Augusto Righi”, Alma Mater Studiorum, Università di Bologna, Via Gobetti 93/2, I-40129 Bologna, Italy
        \and
            INAF, Osservatorio di Astrofisica e Scienza dello Spazio, Via Gobetti 93/3, I-40129 Bologna, Italy\\
            \email{marco.palla@inaf.it}
        \and 
            Institut f\"ur Kernphysik, Technische Universit\"at Darmstadt, Schlossgartenstr. 2, Darmstadt 64289, Germany
             }

   \date{Received 14/03/25; accepted 28/05/2025}

 
  \abstract
   {Europium (Eu) serves as a crucial tracer to understand the origin of rapid neutron-capture process (r-process) elements. For this reason, an extensive effort was made in the last decade to model the chemical evolution of this element in the Galaxy. However, far less attention was reserved to Eu in different galaxies of the Local Group.}
   {By employing detailed and well-tested chemical evolution models, we investigate Eu enrichment across Local Group dwarf spheroidal galaxies, allowing for a direct comparison between model predictions for dwarf galaxies and the Milky Way.}
   {Building upon an r-process enrichment framework that successfully reproduces the observed Eu abundance patterns as well as the supernova and compact binary merger rates in the Milky Way, we build chemical evolution models for the Sagittarius, Fornax, and Sculptor dwarf spheroidal galaxies and test the enrichment scenario against the abundance patterns observed in these galaxies.}
   {Models reproducing the Galactic Eu patterns significantly underestimate the [Eu/Fe] ratios observed in Local Group dwarfs. 
   To address this ``missing Eu'' problem, we estimate the Eu production rate needed to match the observations and explore the potential contributions either from prompt (core-collapse supernova like) or delayed (compact binary mergers) sources, assessing their compatibility with the Milky Way observables.}
   {The same r-process enrichment frameworks cannot reproduce simultaneously the Eu patterns both in the Milky Way and in dwarf galaxies.
   However, a scenario where additional Eu is provided by an increased production from delayed sources at low metallicity can theoretically reconcile the trends observed in the Milky Way and in Local Group dwarfs, 
   because of the small discrepancies ($\lesssim0.1$ dex) between model predictions and observations found in this case.
   Further targeting and modelling of neutron-capture elements in Local Group galaxies are however needed to fill the gaps in our current understanding of the problem.}


   \maketitle
%

\section{Introduction}
\label{s:intro}

The majority of elements beyond the Fe peak are produced by
neutron (n)-capture processes which can be rapid (r-process) or slow
(s-process) with respect to the $\beta$-decay of the nuclei. Understanding
which are the astrophysical sites of these two processes has become
one of the major challenges in stellar physics and galactic chemical evolution.
The main production sites of r-process elements, in particular, are still subject of intense study and debate \citep[see][for a review]{Arconesreview23}, with both core-collapse supernova (CC-SN)-like objects, such as magneto-rotationally driven SNe \citep[hereafter MRD-SNe, see e.g.][]{Winteler12,Nishimura15,Nishimura17,Reichert21,Reichert23}, collapsars \citep[e.g.][and references therein]{Siegel19}, or magnetars \citep[e.g.][]{Patel2025} and merging of compact objects \citep[neutron star-neutron star or neutron star-black hole, hereafter just referred to as MNS, e.g.][]{Argast04,Thielemann17,Cowan21} having been proposed.

The relatively easy detection of absorption lines  of Europium (compared to most of other n-capture elements, see discussion in e.g. \citealt{Hansen2014, Hansen2018, Lombardo2024, AlencastroPuls2025}), which is one of the few n-capture elements almost entirely built by the r-process (\citealt{Burris00,Bisterzo11,Prantzos20}), together with the huge amount of time invested in Galactic observations (also through large spectroscopic surveys, e.g. Gaia-ESO, GALAH), has allowed us to extensively test the effect of these sources on the observed abundances. 
In this way, the concurrent production by CC-SN-like sources (also named ``prompt sources'') and MNS (also named ``delayed sources'' due to their longer timescales, e.g. \citealt{Cote2017,Hotokezaka2018,Simonetti19,Greggio21}) has became a widely accepted scenario (e.g. \citealt{Argast04,Matteucci14,Cescutti15,Cote19,Tsujimoto21,Molero2023MNRAS.523.2974M,Van2023A&A...670A.129V}).\\

The fairly solid theoretical explanation for the Eu abundance patterns in the Milky Way (MW) 
fostered the interest in this element as a potential chemical tag to discriminate between in-situ and accreted populations in the stellar halo of the Galaxy (e.g. \citealt{Matsuno21,Monty24,Ernandes24}), an interest reinforced also by the fact that it appears to be basically unaffected by the chemical processes internal to Globular Clusters (GCs, e.g. \citealt{Roederer11,McKenzie22}).

Beyond the MW, ultra-faint dwarf (UFD) galaxies in the Local Group (LG), with typical metallicities [Fe/H]$\lesssim-2/-3$ dex \citep{Ji16a,Ji16b,Simon23}, provide excellent environments for unraveling the early galactic chemical evolution of the r-process elements, as due to the stochastic nature of the enrichment, namely, the limited number of nucleosynthetic events that enriched their oldest stars (e.g. \citealt{Hartwig19}). 
Indeed, the very large scatters in abundances observed in these galaxies are among the few keys for determining event occurrence rates and yield variations across different r-process sources (e.g. \citealt{Alexander23,Cavallo23}).

On the other side, much less attention has been devoted to the theoretical study of n-capture elements at larger metallicities, i.e. beyond the regime dominated by stochastic enrichment, in more massive satellites, namely in the most massive dwarf spheroidal (dSph) galaxies. 
This happened despite the good knowledge on the star formation and general chemical evolution history of such objects (e.g. \citealt{Lanfranchi04,deBoer12,deBoer15,Viencenzo14}) and the presence of several observational programs also targeting heavy elements (e.g. \citealt{Letarte10,McWilliam13,Lemasle14,Hill19,Reichert20} and references therein).
Works on the theoretical interpretation of Eu abundances in dSph are in fact lacking, with only few studies focusing on the modelling of the evolution of heavy elements in these galaxies (e.g. \citealt{lanfranchi2008,hirai2015,Vincenzo15,Molero21}). 
The situation is even worse when it is matter of directly applying and comparing the modelling framework of r-process nucleosynthesis adopted for the MW to the chemical evolution of these galaxies. \citet{Skula20} studied the abundance patterns of Eu in Sagittarius, Fornax and Sculptor dSph in comparison to the ones in the Galaxy, proposing that the combination of quick/prompt and delayed r-process sources is able to self-consistently explain the chemical abundance pattern observed in the MW and its dwarf satellite galaxies. However, their adopted approach is purely schematic, without the implementation of any theoretical model. In fact, the same authors pointed out the need for follow-up studies including detailed chemical evolution modelling of both the MW and its dwarf satellite galaxies to investigate the r-process enrichment. \\

In this work, we focus on the theoretical interpretation of Eu enrichment in the three most massive dSph galaxies of the LG, namely Sagittarius, Fornax and Sculptor. We use detailed chemical evolution models that adopt the most recent and up-to-date prescriptions explaining the n-capture abundance patterns and measured CC-SN and MNS rates in the MW galaxy (\citealt{Palla20,Molero2023MNRAS.523.2974M}).
In this way, we can probe whether the accepted scenario to explain the Galactic constraints can be also applied to galaxies exhibiting different star formation histories (hereafter, SFHs), or modifications in the r-process enrichment framework are needed. 

It is worth highlighting that this type of analysis is vital in the light of the use of Eu for chemical tagging (e.g. \citealt{Matsuno21,Monty24}). Other chemical elements routinely used for chemical tagging have well established production mechanisms, which are tested on the abundance ratio patterns of different galaxies in diagrams such as the [$\alpha$/Fe] vs. [Fe/H] \citep[e.g.][]{MatteucciBrocato90,Matteucci12} or [Mg/Mn] vs. [Al/Fe] ones (e.g. \citealt{Horta21, Fernandes23}, but see also \citealt{Vasini2024}).
Since only a detailed knowledge of the element production sources allows for a precise characterisation of the evolutionary path within a galaxy, it is crucial to extend the study of Eu to galactic ecosystems with different 
star formation and chemical evolution histories. \\

The paper is organised as follows. 
In Section \ref{s:data}, we present the data used to constrain the models.
In Section \ref{s:model}, we discuss the model assumptions and ingredients, including the SFHs of different galactic systems and the nucleosynthesis prescriptions for n-capture elements. 
In Section \ref{s:std_scenario}, we describe in detail the r-process enrichment scheme that has been successfully tested against the data for the MW and apply it to the dwarf galaxies object of our study.
Due to the systematic Eu underproduction that we find when implementing the Eu enrichment scheme for the MW in models for dwarf galaxies, in Section \ref{s:missing_Eu} and \ref{s:testing_scenarios} we quantify and explore the possible origin of the ``missing Eu'' in these galaxies.
Finally, in Section \ref{s:discussion_conclusion} we discuss our findings and draw our main conclusions.

\section{Observational data}
\label{s:data}

In the following, we provide the details of the datasets adopted throughout this work to trace Eu enrichment in LG galaxies.
Due to the different objects and extended metallicity intervals treated in this study, we lack homogeneous stellar sampling in terms of derivation of chemical abundances. Therefore, we perform a careful collection of the best options available in the literature.

\subsection{MW data}
\label{ss:dataMW}

Despite the plethora of Galactic spectroscopic surveys in the last decade, we still lack a homogeneous sample of stars spanning the full metallicity range with measured Eu abundances, even for the Solar Neighbourhood.
Therefore, in this work we take advantage of abundance data from different surveys to sample the entire metallicity range within the Galaxy. \\

To sample the intermediate/metal-rich range ([Fe/H]$\gtrsim$-1.25 dex), we consider chemical abundances from field stars and open clusters (OCs) in the Gaia-ESO survey (\citealt{Gilmore22,Randich22}). In particular, we select field stars with Galactocentric radii in the range $7< \, R_{\rm GC}$/kpc~$<9$ (therefore sampling the thick and thin disk in the solar region) with spectra obtained with UVES at high resolution, $\mathcal{R}\sim 47\,000$.
Going more into the details, we consider stars from the sample of \citet{Viscasillas2022A&A...660A.135V}, 
with selection also described in \citet{Molero2023MNRAS.523.2974M}.
For the OCs, instead, we adopt the sample of Gaia-ESO OCs from \citet{Magrini23}, comprising OCs older than 100 Myr (to avoid biases in abundance determination, see \citealt{Spina2022Univ....8...87S,Magrini23,Palla24}), selected in Galactocentric radius as above.

To sample the metal-poorer regime, we use both stars from the Measuring at Intermediate metallicity Neutron-Capture Elements (MINCE) survey (\citealt{Cescutti22,Francois24}) and a collection of individual studies (\citealt{Mishenina01,Hansen12,Ishigaki13,Roederer14,Li22}) extracted from the Stellar Abundances for Galactic Archaeology (SAGA) database (hereafter referred as ``SAGA'' stars, \citealt{Suda08}). For the latter, we consider only studies targeting stars at high-spectral resolution ($\mathcal{R}\gtrsim 40\,000$), in order to consider only stars observed at comparable resolution to Gaia-ESO and MINCE (the latter with $\mathcal{R}$ up to $100\,000$).
In order to avoid contamination of stars from accreted structures and consider only stars formed in-situ in the metal-poor MW, we integrate the stellar orbits by adopting the \citet{McMillan17} potential and apply the same selection criteria for in-situ stars as described in \citeauthor{Monty24} (\citeyear{Monty24}). 
Finally, to avoid stars that are most likely polluted by an asymptotic giant branch (AGB) companion and, thus, do not reflect the chemical composition of their birth gas cloud, we select only the stars with [Ba/Eu]~$<$~0 and [Ba/La]~$<$~0.

Applying all the above mentioned selection criteria, we end up with 1264 field stars and 13 OCs from the Gaia-ESO survey, 15 stars from MINCE and 39 ``SAGA'' stars.

\subsection{Dwarf galaxy data}
\label{ss:dataLG}

As mentioned in Section \ref{s:intro}, here we consider the most massive and well observed dSphs in the LG, namely Sagittarius, Fornax and Sculptor. 
For all three galaxies, we select abundances from studies resting on high-resolution spectroscopy that provide Eu abundance determinations for a consistent number of stars ($>30$) spanning an extended metallicity range ($> 1$ dex in [Fe/H]). 
For the full description of the samples, we address the reader to the original papers. However, we provide some details of the adopted datasets below: 

\begin{itemize}
    \item {\bf Sagittarius (Sgr)}: for this galaxy, we adopt the sample presented in \citet{Liberatori25}. The study presents 37 red-giant branch stars observed by means of the FLAMES-UVES high-resolution spectrograph ($\mathcal{R}\sim 47\,000$), all with available Eu abundances. The data cover a large range of metallicities, namely from [Fe/H]$\sim -$2 to $\sim -$0.4 dex, therefore sampling all the main stellar populations of the galaxy.
    It is worth noting that the metal-poor stars in the sample are selected in order to avoid contamination by the metal-poor globular cluster M54, which biases the abundance pattern of the Sgr dSph (see, e.g. \citealt{Minelli23} for a detailed discussion);
    \item {\bf Fornax (For)}: for this dSph, we take advantage of the sample presented in \citet{Reichert20}. The sample is built on a collection of spectra of red-giant branch stars, both from the archive and unpublished, obtained by means of the HIRES, UVES, and FLAMES-GIRAFFE spectrographs (all at resolution $\mathcal{R}\gtrsim 20\,000$). The spectra adopted in this study were processed  in order to obtain homogeneous abundance measurements for all the targets (see \citealt{Reichert20} for the details). For For dSph, 128 stars are available, of which 108 with available Eu abundances, across the metallicity range $-1.6\le$~[Fe/H]/dex~$\le -0.4$; 
    \item {\bf Sculptor (Scl)}: here, we exploit the study by \citet{Hill19}. In the study, 99 red-giant branch stars in the center of the Scl dSph were observed through FLAMES-GIRAFFE and FLAMES-UVES (all at resolution $\mathcal{R}\gtrsim 20\,000$), for which a homogeneous abundance analysis was performed. Of these stars, 51 have available Eu abundances, across a metallicity distribution spanning $-2.1 \lesssim$ [Fe/H]/dex $\lesssim -0.9$.\\
\end{itemize}

\begin{figure}
    \centering
    \includegraphics[width=0.95\columnwidth]{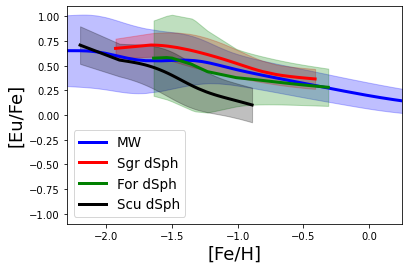}\\
    \includegraphics[width=0.95\columnwidth]{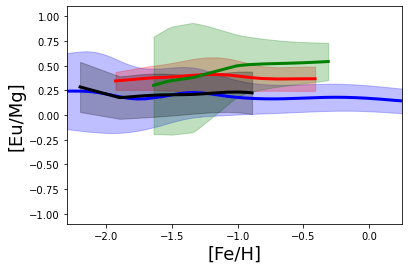}
    \caption{[Eu/Fe] (upper panel) and [Eu/Mg] (lower panel) vs. [Fe/H] trends in the Local Group. The solid lines are non-parametric Gaussian KDE regressions computed for each galaxy basing on the datasets adopted 
    throughout the paper. The shaded areas represent the $1\sigma$ confidence interval of the regressions.}
    \label{fig:trends_LG}
\end{figure}

The trends in [Eu/Fe] and [Eu/Mg] vs. [Fe/H] for each of the galaxies considered in this work are shown in Fig. \ref{fig:trends_LG}. 
The Figure shows a different behavior of the chemical pattern both depending on the galaxy and the abundance ratio we are looking at, namely [Eu/Fe] (upper panel) and [Eu/Mg] (lower panel). 
In particular, we clearly see that while Sgr and For show a similar trend in [Eu/Fe] in comparison with the MW, the same does not happen for [Eu/Mg], where the galaxies generally have higher abundances of [Eu/Mg], with a stark overabundance especially in Fornax at higher metallicities. 
On the other hand, Scl shows much lower [Eu/Fe] ratios relative to those observed in the Galaxy at higher metallicities, but at the same time an [Eu/Mg] superposed to that of the Galaxy in the metallicity interval in common.
Despite in this work we focus on the abundance ratios trends as function of [Fe/H], it is worth noting that even considering [Mg/H] on the x-axis, similar relative behaviour in [Eu/Mg] between different galaxies is also found. In turn, this gives further confirmation of what is observed in Fig. \ref{fig:trends_LG}.

In this work we try to explain the above mentioned differences by using a detailed model of chemical evolution, considering the peculiar SFHs of the different galaxies and implementing different sources of Eu enrichment (prompt and delayed, see also Section \ref{s:model}). 
In particular, we test whether the different SFHs alone can explain the differences in these trends, starting from models that exploit all the available constraints for r-process production (yields and event rates) and well reproduce the abundance patterns observed in our Galaxy.\\

It is it worth mentioning that in Fig. \ref{fig:trends_LG} and throughout this work, all the abundances are rescaled to the solar ones by \citet{Grevese98}. 
In spite not guaranteeing a fully uniform comparison, possible only with fully homogeneous datasets, this assures a first order of consistency, which is more than adequate for the purposes of this work.

\section{Chemical evolution of r-process elements}
\label{s:model}

In this Section, we present the main prescriptions adopted to model the chemical evolution of r-process elements, and especially Eu, in different systems. 
In particular, in Section \ref{ss:chem_evo} we briefly summarize the main features of the 
chemical evolution models adopted for the solar vicinity and LG dwarf galaxies, whereas in Section \ref{ss:rprocess} we focus on the prescriptions (e.g. stellar yields, delay-time-distributions) adopted for the enrichment sources of r-process elements.

\subsection{Chemical evolution models}
\label{ss:chem_evo}

All the chemical evolution models discussed in this work rest on the following basic equation that describes the chemical evolution of a given element $i$ (see, e.g. \citealt{Matteucci12}):
\begin{equation}
    \Dot{G}_i (t) = -\psi(t)\, X_i(t)\, + \,R_i(t)\, +\, \Dot{G}_{i,inf}(t)\, -\, \Dot{G}_{i,out},
    \label{eq:basic_chemevo}
\end{equation}
where $G_i(t) = X_i(t)\,G(t)$ is the fraction of the gas mass in the form of an element $i$ and $G(t)$ is the fractional gas mass.
$X_i(t)$ represents the abundance fraction in mass of a given element $i$, with the summation over all elements in the gas mixture being equal to unity.

The first term on the right-hand side of Eq. \eqref{eq:basic_chemevo} corresponds to the rate at which an element $i$ is removed from the ISM due to star formation. The star formation rate ($\psi(t)$, hereafter SFR) is parametrised according to the Schmidt-Kennicutt law (\citealt{Kennicutt98}), with the star formation efficiency (SFE) $\nu$ as the control parameter that represents the SFR per unit mass of gas.

$R_i(t)$ (see, e.g. \citealt{Palla20b} for the complete expression) takes into account the nucleosynthesis from different stellar sources, weighted according to the initial mass function (IMF). This term also includes the products originating from binary systems such as MNS (see Section \ref{ss:rprocess}) and Type Ia SNe.
For the latter, we assume the delay-time-distribution (DTD) by \citet{MatteucciRecchi01}, which enables us to obtain abundance patterns that are similar to those obtained with other literature Type Ia SNe DTDs (see, e.g. \citealt{Matteucci09,Palla21} for details).

The last two terms of Eq. \eqref{eq:basic_chemevo} refer to the gas flows history of the galaxy, namely inflows and outflows, for which different prescriptions are adopted depending on the specific history of star formation of the galaxies under scrutiny (see Sections \ref{sss:MWmodel}, \ref{sss:dSphmodel}). 
In general, we model the rate of gas inflow in terms of an exponentially-decaying law (see, e.g. \citealt{Romano10,Palla20b,Koba20SNIa}):
\begin{equation}
    \Dot{G}_{i,inf} = C_{inf}\, X_{i,inf}\, e^{-t/\tau_{inf}}, 
    \label{eq:infall}
\end{equation}
where $C_{inf}$ is the normalization constant, which is set to reproduce the total gas mass accreted by the infall, $X_{i,inf}$ is the chemical abundance of the element $i$ of the infalling gas, here assumed to be primordial, and $\tau_{inf}$ the infall timescale. For what concerns galactic winds, we assume them to be proportional to the SFR (e.g. \citealt{Vincenzo15,Molero21,Palla24b}):
\begin{equation}
    \Dot{G}_{i,out} = \omega_i \, \psi(t),
    \label{eq:outflow}
\end{equation}
where $\omega_i$ is the mass loading factor free parameter, assumed to be the same for all chemical elements $i$. For the MW, $\omega_i = 0$ (see \citealt{Meioli09,Spitoni09,Hopkins23}).\\

All the galaxy models adopted in this work adopt the same IMF and nucleosynthesis prescriptions.
For the IMF, we use the one by \citet{Kroupa93}. 
The galaxy-wide IMF (gwIMF) of dwarf galaxies might be different from that derived from local star counts \citep[see, e.g.][and references therein]{Jerabkova2018}. Hence, we explored alternative IMF formulations that, however, do not help to solve the Eu problem. Therefore, we do not discuss the issue of gwIMF variations further.
Concerning the stellar yields, we adopt those by \citet{Karakas10} for low- and intermediate-mass stars, \citet{Nomoto13} for massive stars, and \citet{Iwa99} for Type Ia SNe. 
For the sources of production of n-capture elements, we address the reader to Section \ref{ss:rprocess}.

\subsubsection{MW galaxy}
\label{sss:MWmodel}

To model the evolution of the solar vicinity, we adopt the best model of \citet{Palla20} that has been already implemented to study the evolution of n-capture elements in the MW disk by \citet{Molero2023MNRAS.523.2974M}. The model is a revised version of the two-infall paradigm \citep[][see also \citealt{Spitoni19}]{Chiappini1997,Chiappini2001}, in which two consecutive gas accretion episodes, separated by an age gap of 3.25 Gyr, form the so-called high-$\alpha$ and low-$\alpha$ sequences observed in the Galactic disk.

In this scenario, the first infall takes place on a relatively short timescale ($\tau_{inf} =$ 1 Gyr) and high SFE, ($\nu =$ 2 Gyr$^{-1}$), while the second gas-infall episode happens on longer timescales ($\tau_{inf}\simeq$ 7 Gyr) and lower SFE ($\nu =$ 1.3 Gyr$^{-1}$), in accordance with what claimed by several studies for the solar vicinity (e.g. \citealt{Spitoni20,Palla22,Romano21}, see also, e.g. \citealt{Marcon2010,Koba20SNIa,Koba20} for similar consideration on the Galactic thin disk modelling). 
Indeed, these assumptions allowed to reproduce large survey abundance data (\citealt{Palla20,Spitoni21}), as well as abundance-age diagrams (\citealt{Spitoni19,Spitoni20}) in the Solar Neighbourhood, with also a good agreement with present-day rates of star formation and SNe (see, e.g. \citealt{Molero2023MNRAS.523.2974M} and references therein).\\

\begin{figure}
    \centering
    \includegraphics[width=0.98\columnwidth]{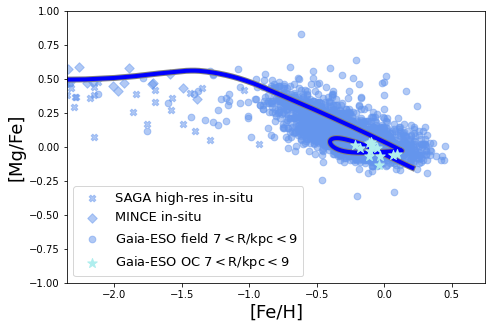}
    \caption{[Mg/Fe] vs. [Fe/H] for the MW. Data are from high-resolution SAGA data selection (see Section \ref{ss:dataMW}, light blue crosses), MINCE (\citealt{Cescutti22}, light blue diamonds), Gaia-ESO field stars (light blue points, \citealt{Viscasillas2022A&A...660A.135V}) and Gaia-ESO open clusters (cyan stars, \citealt{Magrini23}).}
    \label{fig:MW_MgFe}
\end{figure}

The reproduction of the typical [$\alpha$/Fe] abundance pattern in the solar vicinity can be observed in Fig. \ref{fig:MW_MgFe}, where we compare the predictions of the two-infall model in the [Mg/Fe] vs. [Fe/H] diagram with abundances from the data samples described in Section \ref{ss:dataMW}. 
In the Figure, we clearly see the typical loop/ribbon feature as due to the delayed gas accretion episode forging the low-$\alpha$ sequence (see, e.g. \citealt{Spitoni19}), well in agreement with the bulk of MW disk stars and OCs (the latter tracing stars younger than $\sim$5 Gyr) in the solar region.
The comparison with low-metallicity, in-situ stars also assure us that we neither significantly underestimate or overestimate the level of the [$\alpha$/Fe] plateau, therefore preventing any initial substantial bias when we are looking to elements which are much more uncertain in terms of their nucleosynthetic origin.

\subsubsection{Local group dSphs}
\label{sss:dSphmodel}

To model the chemical evolution of LG dSph, we follow previous chemical evolution works by \citet{Molero21} for Fornax and Sculptor and \citet{Mucciarelli17,Minelli23} for Sagittarius. 
Below, we summarize the main model assumptions for the three galaxies:


\begin{itemize}
    \item {\bf Sagittarius (Sgr)}:  we assume a rather long lasting, single episode of star formation for this galaxy, from $\sim$~14 to 6~Gyr ago, after which the Sgr galaxy stops forming new stars, in agreement with results by \citet{Dolphin02} and \citet[but see also \citealt{Siegel07}]{deBoer15}. We assume an infall timescale of $\tau_{inf}$ = 3 Gyr and a SFE of $\nu =$ 0.2 Gyr$^{-1}$ from $\sim$14 to 7 Gyr ago, after which the star-formation rate drops abruptly ($\nu =$ 0.001 Gyr$^{-1}$). This is largely due to gas stripping (for which we assume a mass loading factor $\omega=2$) caused by the interaction with the Galaxy \citep[see][for details]{Minelli23}, starting 7.5~Gyr ago in the model.
    The tidal stripping  also impacts the dSph stellar populations, forming the Sgr stream we observe at the present-day (e.g. \citealt{Monaco07,deBoer15}). Because the adopted dataset for Sgr samples the Sgr main body (see \citealt{Minelli23}), we remove 50\% of the stars with [Fe/H]$< -0.6$ dex in accordance with \citet[and references therein, see also \citealt{Minelli23}]{LawMajewski16}, who showed that tidal stripping removes preferentially metal-poor stars from the Sgr main body;

    \item {\bf Fornax (For)}:  for this galaxy we take into consideration the SFH derived by \citet{deBoer12} from color-magnitude diagram (CMD) fitting analysis, according to which Fornax formed stars at all ages, from as old as 14 Gyr to as young as 0.25 Gyr. 
    For this reason, we model a continuous star formation, characterized by one single long episode lasting 14 Gyr, with infall timescale and SFE equal to $\tau_{inf}=2.25$ Gyr and $\nu = 0.125$ Gyr$^{-1}$, respectively, also assuming galactic winds with mass loading factor $\omega=1$ \citep{Molero21}.
    With such a SFH, the model predicts a present-day stellar mass of $M_* = 3.1 \cdot 10^7$ M$_\odot$, very similar to the one estimated by \citet{deBoer12}, $M_* = 4.3 \cdot 10^7$ M$_\odot$;

    \item {\bf Sculptor (Scl)}: as for the SFH of Scl, we rest on the CMD fitting analysis by \citet{Bettinelli19}. These authors show that most of Scl stars are formed in a single episode of star formation ending $\sim$9-10 Gyr ago.
    We model this SFH with an initial burst ($\nu=$ 0.1~Gyr$^{-1}$) driven by rapid infall of gas ($\tau_{inf}$ = 0.5 Gyr) and a subsequent decline driven by galactic winds, with mass loading factor $\omega=2.5$.
    In this way, the predicted final stellar mass is $M_* = 2.6 \cdot 10^6$ M$_\odot$, in agreement with the observed one, $M_* = 2.24 \cdot 10^6$ M$_\odot$ (\citealt{Bettinelli19}, see also \citealt{deBoer12}).\\
    
\end{itemize}

\begin{figure*}
    \centering
    \includegraphics[width=0.95\columnwidth]{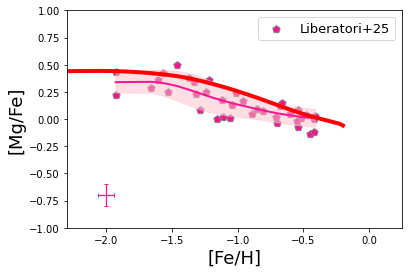}
    \includegraphics[width=0.95\columnwidth]{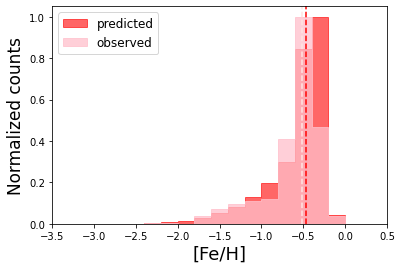}\\
    \includegraphics[width=0.95\columnwidth]{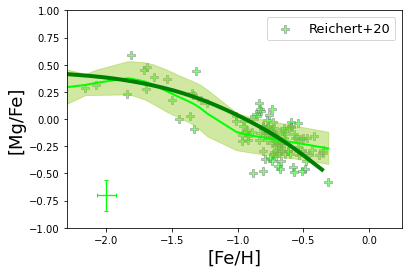}
    \includegraphics[width=0.95\columnwidth]{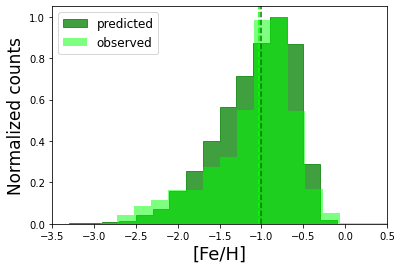}\\
    \includegraphics[width=0.95\columnwidth]{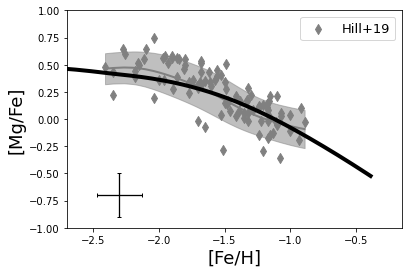}
    \includegraphics[width=0.95\columnwidth]{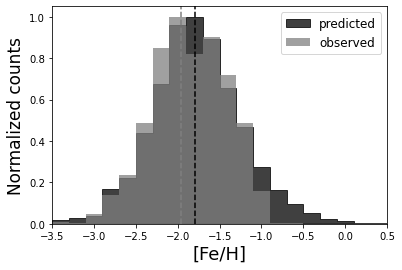}
    \caption{
    [Mg/Fe] vs. [Fe/H] (left panels) and MDF (right panels) for Sagittarius (top panels) Fornax (central panels) and Sculptor (bottom panels). 
    Data for [Mg/Fe] vs. [Fe/H] are from \citet{Liberatori25} for Sagittarius, \citet{Reichert20} for Fornax and \citet{Hill19} for Sculptor. The thin solid lines and shaded areas represent the non-parametric Gaussian KDE data regressions and their 1$\sigma$ confidence interval for [Mg/Fe] vs. [Fe/H]. The thick solid lines are the model predictions. For sake of comparison with observations, in this and the following Figures we show theoretical [X/Fe] vs. [Fe/H] curves for the first 99\% of the predicted cumulative distribution of stars.
    Observed MDFs are from \citet{Minelli23} for Sagittarius and \citet{Molero21} for Fornax and Sculptor.}
    \label{fig:dwarf_model}
\end{figure*}

The above theoretical prescriptions allow us to reproduce the observations available for the different systems. 
This is shown in Fig. \ref{fig:dwarf_model}, where we compare observed and predicted [Mg/Fe] vs. [Fe/H] abundance patterns and metallicity distribution functions (MDFs) for Sgr (top panels), For (central panels) and Scl (bottom panels) dSphs.
The observed MDFs are from \cite{Minelli23} for Sgr and \cite{Molero21} for For and Scl; more details can be found in the original papers.

In general, we observe that the set of parameters adopted for the three systems reproduce well the data trends in the [Mg/Fe] vs. [Fe/H] diagrams (Fig. \ref{fig:dwarf_model} left panels), also considering the observational uncertainties and the non uniform sampling in metallicity (especially for For).
The agreement between observed and modelled abundance patterns demonstrates the suitability of our adopted single-episode SFHs to explain the evolution of these dSphs, without advocating to more complex SFH (an therefore, abundance patterns) as for the two-infall scenario for the MW.
A remarkable agreement is also obtained between the predicted MDFs and the measured ones (Fig. \ref{fig:dwarf_model} right panels). This is further confirmed by looking at the predicted and observed median distributions: the two are almost overlapping in all the three panels, showing a maximum deviation of $\simeq 0.15$ dex for Scl.\\

Despite the large uncertainties in the SFH measurements of the three LG galaxies (e.g. \citealt{Siegel07,deBoer12,deBoer15}) and the theoretical limitations of our pure chemical models in tracking the complex chemo-dynamical evolution of the galaxies, it is important to stress that the simultaneous reproduction of both [$\alpha$/Fe] vs. [Fe/H] abundance patterns and MDFs allows us to place stringent constraints on the evolutionary paths of these systems (see, e.g. the discussion in \citealt{Romano15}). 
Therefore, the above mentioned limitation are not expected to strongly affect our conclusions regarding the evolution of n-capture elements in these galaxies. Indeed, the understanding of the nucleosynthetic origin of the elements beyond the Fe-peak is very limited, making the uncertainties on elemental nucleosynthesis much more relevant than the ones arising from the SFHs of the galaxies.

\subsection{r-process elements nucleosynthesis}
\label{ss:rprocess}

As explained in Section \ref{ss:chem_evo}, all the models adopt the same nucleosynthesis prescriptions for n-capture elements. 
As Eu is the main topic of this work and it is a typical r-process element (produced at $\sim97\%$ by the r-process at solar system formation, \citealt{Burris00}), here we focus on r-process nucleosynthesis only.\\

We consider two channels for the production of r-process elements: MNS and MRD-SNe. MNS are computed as systems of two neutron stars of 1.4 M$_\odot$ with progenitors in the 9–50 M$_\odot$ mass range. 
Their rate is computed as the convolution between a given DTD and the SFR \citep[see][]{Molero21,Molero2023MNRAS.523.2974M}: 
\begin{equation}
    R_{MNS}(t)=k_\alpha\, \int_{\tau_i}^{min(t,\tau_x)} \alpha_{MNS} \, \psi(t-\tau) \, f_{MNS}(\tau) \,d\tau
    \label{eq:MNS_rate}
\end{equation}
where $k_\alpha$ is the number of neutron-star progenitors per unit mass in a stellar generation ($0.0041$ for a \citealt{Kroupa93} IMF), $\alpha_{MNS}$ is the fraction of stars in the correct mass range which can give rise to a double neutron-star merging event, set to $2\cdot10^{-3}$ as reference (\citealt{Molero2023MNRAS.523.2974M}), and $f_{MNS}$ is the DTD of MNS. 
For the DTD, we adopt the formulation of \citet[see also \citealt{Greggio21}]{Simonetti19}, where the nuclear lifetime of the secondary progenitor component accounts for an initial plateau lasting up to 40 Myr. This is followed by a power-law decay proportional to $t^{-1}$, a form commonly used in chemical evolution studies (e.g., \citealp{Cote19, Cavallo22}), which accounts for systems where the delay is primarily driven by gravitational radiation.

The yields of r-process elements from MNS have been obtained by scaling to solar the yield of Sr measured in the re-analysis of the spectra of the kilonova AT2017gfo by \citet[after having considered uncertainties in
its derivation; see \citealt{Molero21} for details]{Watson19}. The Eu yield obtained in this way is $Y_{\mathrm{Eu}}(MNS)=3\cdot10^{-6}$ M$_\odot$ (per merging event), also consistent with the theoretical calculation of \citet{Korobkin12} and with estimates from \citet{Matteucci14}.\\

For what concerns MRD-SNe, here we assume that they originate from a fraction of stars in the mass interval between 10 and 80 M$_\odot$. This fraction is regulated by the $\alpha_{MRD}$ parameter, which here is set to 0.2 as in \citet{Molero2023MNRAS.523.2974M}, fine-tuned to reproduce the r-process pattern in the MW disk.

The set of yields adopted for the MRD-SNe is the model L0.75 of \citet{Nishimura17}, also chosen to be consistent with the best-fitting model of \citet{Molero2023MNRAS.523.2974M} for the MW disk.

\section{A well-behaved model for the MW}
\label{s:std_scenario}

In this Section, we show the predicted behaviour of Eu enrichment for the galaxies object of our study under the scenario that best reproduces the [Eu/Fe] vs. [Fe/H] average trend in the solar vicinity. 
The parameters concerning the r-process nucleosynthesis, namely the fraction of progenitors (for MNS and MRD-SNe) in the right mass interval 
and the yields, are listed in Tab. \ref{tab:params_rproc}. 

\begin{table}
    \centering
    \caption{Main parameters for r-process enrichment in the calibrated MW disk model (\citealt{Molero2023MNRAS.523.2974M}).}
    {\setlength{\extrarowheight}{3pt}
    \begin{tabular}{c c c c}
        \hline
        $\alpha_{MNS}$ & $Y_{\rm Eu}(MNS)$ $^{(1)}$ & $\alpha_{MRD}$ & $Y_{\rm Eu}(MRD)$ $^{(2)}$\\[0.1cm]
        \hline
        \hspace{0.15cm} $2\cdot 10^{-3}$ \hspace{0.15cm} & \hspace{0.15cm} $3\cdot10^{-6}$ M$_\odot$ \hspace{0.15cm} & \hspace{0.15cm} 0.2 \hspace{0.15cm} & \hspace{0.15cm} $4.69\cdot10^{-7}$ M$_\odot$ \hspace{0.15cm}\\[0.1cm]
        \hline
    \end{tabular}
    }\\[0.15cm]
    \footnotesize{{\bf Refs}: $^{(1)}$ \citet{Molero21}, $^{(2)}$ \citet{Nishimura17}}
    \label{tab:params_rproc}
\end{table}

\begin{figure}
    \centering
    \includegraphics[width=0.95\columnwidth]{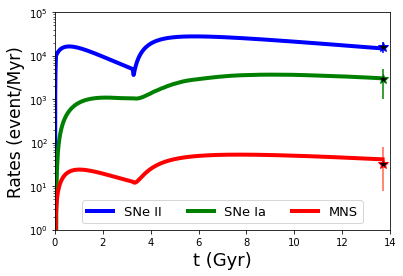}
    \caption{Predicted evolution of CC-SN, Type Ia SN and MNS rates in the MW compared with present-day observations from \citet[CC-SNe]{Rozwado21}, \citet[Type Ia SNe]{Cappellaro97} and estimate from \citet[MNS]{Abbott21}, respectively.} 
    \label{fig:rates_MW}
\end{figure}

Having fixed the DTD function (see \ref{ss:rprocess}) and the SFR behavior for the MW (from the two-infall model, see \ref{sss:MWmodel}), the adopted fraction of systems originating MNS ($\alpha_{MNS}$) allows us to well reproduce the measured rate of MNS in the Galaxy. 
This is shown in Fig. \ref{fig:rates_MW}, where we report the predictions of the MW model for the rates of CC-SN, Type Ia SN and MNS compared to the observations (all averaged over the whole disc).
It is worth noting that for MNS, we consider the cosmic rate observed
by \citet{Abbott21}, i.e. $320^{+490}_{-240}$ Gpc$^{-3}$ yr$^{-1}$, converting it to a Galactic rate (see Section 5.2 in \citealt{Simonetti19} for details on the derivation). The obtained rate for the MW ($32^{+49}_{-24}$ Myr$^{-1}$) is also in agreement  within the error bars with the rate of \citet{Kalogera04} from binary pulsars.

\begin{figure}
    \centering
    \includegraphics[width=0.95\columnwidth]{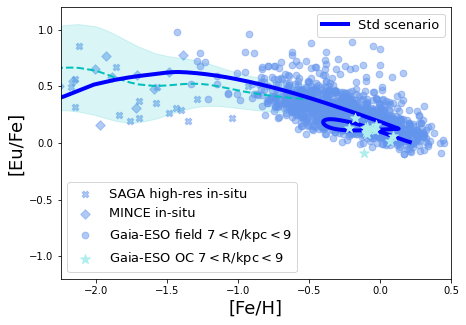}
    \caption{[Eu/Fe] vs. [Fe/H] for the solar vicinity. Data are a selection from the SAGA database (light blue crosses, see Section \ref{ss:dataMW}), MINCE (light blue diamonds, \citealt{Francois24}), Gaia-ESO field stars (light blue points, \citealt{Viscasillas2022A&A...660A.135V}) and Gaia-ESO open clusters (cyan stars, \citealt{Magrini23}). 
    The cyan thin dashed line and shaded area represent the non-parametric Gaussian KDE data regressions and its 1$\sigma$ confidence interval. The thick solid blue line are the predictions of the standard model for the MW (see text).}
    \label{fig:std_MW}
\end{figure}

In Fig. \ref{fig:std_MW}, we show instead the predictions for the [Eu/Fe] vs. [Fe/H] trend in the solar vicinity, as compared to the data for the MW described in Section \ref{s:data}. 
It is worth noting that in this figure (as well as the following ones), we limit the metallicity to values greater than [Fe/H]~$=-$2.5 dex. This is because we are not interested in the scatter caused by the stochastic enrichment of Eu that characterizes the very metal-poor regime (e.g. \citealt{Cescutti15,Wanajo21}, see also Section \ref{s:intro}). 
Rather, the goal of this work is to study and reproduce the trend of stars at metallicities [Fe/H]$\gtrsim -2$ dex, namely at metallicities where the chemical mixing starts to be effective (e.g. \citealt{Schoenrich19}) and the scatter is more likely to be dominated by measurement errors (see, e.g. \citealt{Molero2023MNRAS.523.2974M}).  \\

Referring back to Fig. \ref{fig:std_MW}, we stress that the r-process enrichment scenario adopted here well agrees with the data trend over the entire metallicity range of interest, from low metallicities ([Fe/H]$\sim-$2 dex) up to the metallicities observed in Galactic disk field stars and OCs. In particular, the latter trace the enrichment in the Galactic disk in the last few Gyrs, and they are very well reproduced by the typical loop/ribbon feature caused by the delayed second infall episode (see also Fig. \ref{fig:MW_MgFe} and \citealt{Spitoni19,Palla20,Molero2023MNRAS.523.2974M}). 

The simultaneous reproduction of Galactic rates and chemical abundance patterns confirm the results of previous studies who claimed that both prompt and delayed Eu sources are needed to reproduce the observations (e.g. \citealt{Matteucci14,Cescutti15,Cote19,Tsujimoto21,Molero2023MNRAS.523.2974M,Van2023A&A...670A.129V}). 
However, it is worth noting that the prompt source for Eu enrichment, here represented by MRD-SNe (alternative sources are collapsars, e.g. \citealt{Siegel19} and references therein), dominates over the enrichment by MNS (see also \citealt{Van2023A&A...670A.129V}).
Indeed, by computing the present mass fraction of Eu in the ISM produced by individual sources, we get $X({\rm Eu})_{MNS}= 5.6 \cdot 10^{-11}$ and $X({\rm Eu})_{MRD}= 5.2 \cdot 10^{-10}$, namely only $\simeq 10 \%$ of the total Eu budget is produced by delayed sources such a MNS.
In turn, this indicates a limited impact of MNS in shaping the observed [Eu/Fe] abundance patterns in the MW 
(see also \citealt{Molero2023MNRAS.523.2974M}).

\subsection{Applying the scenario to LG dSphs}
\label{ss:std_scenario_dsph}

We now apply the prescriptions for r-process enrichment that explain the Galactic rates and Eu abundance patterns to the dwarf galaxies that are the subject of this study.
Such an approach was already adopted to the modelling of r-process in dwarfs \citep[e.g.][]{Vincenzo15,Molero21}, and stems from the fact that, at variance with the MW, no MNS rate estimation can be inferred directly / derived in a straightforward manner \citep{Kalogera04,Simonetti19} for these systems.

\begin{figure}
    \centering
    \includegraphics[width=0.95\columnwidth]{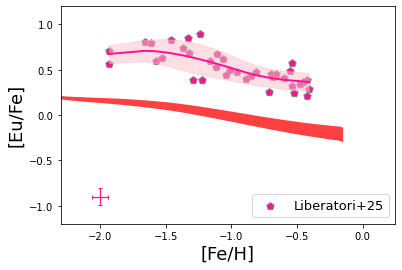}\\
    \includegraphics[width=0.95\columnwidth]{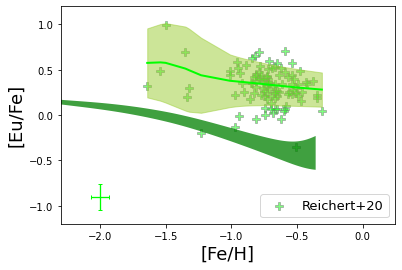}\\
    \includegraphics[width=0.95\columnwidth]{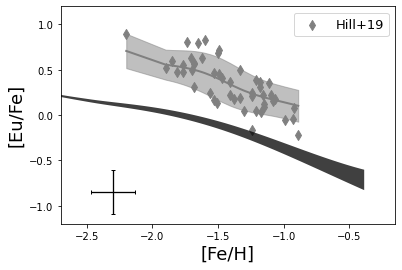}\\
    \caption{[Eu/Fe] vs. [Fe/H] for Sagittarius (top panel), Fornax (central panel) and Sculptor (bottom panel). 
    Data for are from \citet{Liberatori25} for Sagittarius, \citet{Reichert20} for Fornax and \citet{Hill19} for Sculptor. 
    The dark shaded areas are the range of predicted values adopting the reference MNS yield and a factor 5 larger MNS yield to account for the uncertainties in MNS yield measurements by \citet{Watson19}.    
    The thin solid lines and light shaded areas represent the non-parametric Gaussian KDE data regressions and their 1$\sigma$ confidence interval for [Eu/Fe] vs. [Fe/H].
    }
    \label{fig:std_dwarfs}
\end{figure}

The obtained [Eu/Fe] vs. [Fe/H] abundance patterns for Sgr, For and Scl dSphs are shown in Fig. \ref{fig:std_dwarfs} (top, central and bottom panels, respectively). In all the panels, we show the range of predicted values adopting the reference MNS yield (see Tab. \ref{tab:params_rproc}) and a factor 5 larger MNS yield, i.e. within the uncertainties in MNS yield measurements by \citet[see \citealt{Molero21}]{Watson19}.
It is evident that the three models show a marked deficiency in Eu production with respect to what is observed, with a typical underestimation of [Eu/Fe] relative to the data regression by $\sim 0.5$ dex in all systems. Even a factor 5 increase in the MNS yield does not help in  reducing significantly the large displacement between the observed and predicted [Eu/Fe] trends.\\

Therefore, the r-process enrichment scenario that best reproduces the measurements in the solar vicinity has severe limitations when dealing with other galactic systems. 
It is worth stressing that such inconsistency between Galactic and extra-Galactic trends has not been pointed out before (see, e.g. \citealt{Skula20,Molero21,Ernandes24}). 
Earlier works studying the MW galaxy were basing on previous MNS rate measurements (\citealt{Kalogera04,Abbott17}, see e.g. \citealt{Simonetti19,Grisoni20}).
Indeed, the present-day MNS rate by \citet{Abbott17} provides order of magnitude larger values relative to more up to date measurements ($R_{MNS}= 154^{+320}_{-122}$ Myr$^{-1}$ instead of $R_{MNS}= 32^{+49}_{-24}$ Myr$^{-1}$). This means that larger $\alpha_{MNS}$ (of the order of $\gtrsim 10^{-2}$) have to be adopted in the models for the MW to fit the present-day Galactic MNS rate.
Therefore, previous studies on dwarf galaxies that apply \(\alpha_{MNS}\) as inferred to reproduce the earlier MNS rate in the MW (e.g. \citealt{Molero21}) achieve a satisfactory agreement with the observed abundances. However, as stated above, a reassessment of the observed MNS rate necessitates reducing the probability of MNS events (i.e. reducing $\alpha_{MNS}$) in the model, ultimately making MNS an inefficient channel for r-process production.

In turn, an increase in the MNS contribution to Eu enrichment by increasing the $\alpha_{MNS}$ parameter should alleviate the tension between models, as changes to delayed sources contribution potentially have stronger effects in the abundance ratios of dwarf galaxies than more massive ones, due to the time-delay model (see \citealt{Matteucci12,Matteucci21}). Indeed, the slower rate of metal enrichment would allow for more Eu accumulation by delayed sources at low-metallicities, boosting the [Eu/Fe] to larger values. 
However, just increasing the $\alpha_{MNS}$ parameter across the whole time evolution is not a viable solution, as it would cause an overestimation of the present-day MNS rate in the Galaxy measured by \cite{Abbott21}.


\section{The missing Eu in LG dSph}
\label{s:missing_Eu}

The model for the dwarfs of the LG show a clear lack of Eu, as illustrated in Fig. \ref{fig:std_dwarfs}. In this section, we try to estimate the missing rate of Eu production of our model within the three dwarf galaxies analysed in this work. \\

Without making any hypothesis on the origin of this lack of r-process element production, we run a grid of models where we add a fictitious source producing Eu up to a [Fe/H] threshold (hereafter, [Fe/H]$_{\rm thresh}$): 
\begin{equation}
    {\rm "Missing\, Eu"\, rate} = \left\{
    \begin{array}{ll}
 x\hspace{1cm} {\rm if} \, \, {\rm [Fe/H]}\leq {\rm [Fe/H]_{thresh}}\\[0.1cm]
 0 \hspace{1cm} {\rm if} \, \, {\rm [Fe/H]} > {\rm [Fe/H]_{thresh}}\\
\end{array}, \right.
\label{eq:missingEu}
\end{equation}
where $x$ is the "missing Eu" production rate normalised to the SFR. For the gridding scheme, we adopt an array with respective spacing of 0.05 dex for [Fe/H]$_{\rm thresh}$ and 5$\cdot 10^{-11}$ for the "missing Eu" $x$.
We model this additional production rate per unit SFR, in order to compare directly the values we find in the different galaxies we model. 
For the grid of models, we compute their likelihood relative to the observed Eu abundance trends for each individual galaxy (Sgr, For and Scl), as well as the total weighted likelihood for the three galaxies together (see Appendix \ref{a:Max_Like} for the details of the calculation).

The approach described above may appear simplistic, as the missing production rate of r-process may vary depending on metallicity (or time) with a more complex shape than the heaviside step function adopted in Eq. \eqref{eq:missingEu} (see, e.g. Fig. 2 of \citealt{Cote19}). However, 
some arguments can be put forward to justify our choice, for instance: i) the data samples adopted for the different dSph are limited in size ($\lesssim 100$ stars per galaxy): this is at variance with typical samplesfor the MW, which can count on thousands of stellar sources, allowing more sophisticated statistical analysis. ii) Eu abundances are measured for stars in a restricted metallicity range (i.e., it is not possible to probe Eu evolution across the whole metallicity range as inferred from the galaxy's MDF): therefore, it is difficult to properly sample the missing Eu budget in different metallicity intervals.  
It is also worth saying that the main goal of this analysis is to give a first estimate of this missing Eu budget. In fact, this is the first time that the problem is illustrated from the theoretical point of view. Therefore, it is important to give a first estimate, before proceeding with a more in-depth analysis.

\begin{figure}
    \centering
    \includegraphics[width=0.99\columnwidth]{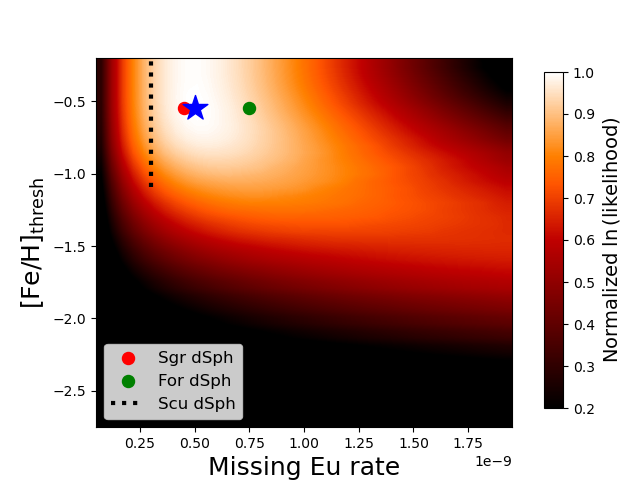}
    \caption{Total weighted likelihood for the sample dwarf galaxies as function of "missing Eu" rate (normalised to the SFR) and  metallicity threshold [Fe/H]$_{\rm thresh}$. The model setup with maximum total weighted likelihood is indicated with the blue star.  
    Setups with maximum likelihood for Sagittarius, Fornax and Sculptor are indicated with a red point, a green point and a dashed line, respectively.}
    \label{fig:missEu_likel}
\end{figure}

The results of the likelihood computation are shown in Fig. \ref{fig:missEu_likel}.  In particular, we show the colormap for the total weighted likelihood, together with the models obtaining the maximum likelihood for the individual galaxies. 
The "missing Eu" rate needed to reconcile the observed [Eu/Fe] trends with the model predictions is similar for the different dwarf galaxies (within a factor of $\sim$2 difference, see Fig. \ref{fig:missEu_likel} symbols). 
The best models for individual galaxies are also in line with the maximum total weighted likelihood for the sample, giving a "missing Eu" rate (normalised to the SFR) of $5 \cdot 10^{-10}$.
The analysis also highlights that this additional Eu production must be in place up to a metallicity of [Fe/H]=$-$0.55 dex. 
Notice that the Scl stellar distribution extends only up to [Fe/H]$\sim-1$ dex (see Fig. \ref{fig:dwarf_model}) and this is why the likelihood on the metallicity threshold axis for this galaxy is not reliable above this metallicity. This is why we find that Scl models with [Fe/H]$_{\rm thresh}\geq-1.05$ dex have the same likelihood for a given "missing Eu" rate.\\

\begin{figure}
    \centering
    \includegraphics[width=0.95\columnwidth]{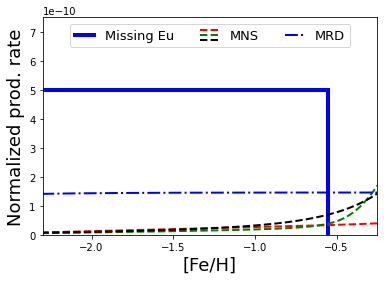}
    \caption{Eu production rate normalised to the SFR by different sources ("missing Eu": thick solid line, MNS: dashed lines, MRD-SNe: dash-dotted line) in the models for LG dSphs with maximum total weighted likelihood (blue star in Fig. \ref{fig:missEu_likel}).
    Different Eu production rates by MNS are found for different galaxies (black: Scl, green: For, red: Sgr) due to their different SFHs (see text).}
    \label{fig:missEu}
\end{figure}

In summary, we can say that in LG dwarfs up to metallicities [Fe/H]~$\sim -$0.5 dex an additional Eu production rate (normalised to the SFR) of $5 \cdot 10^{-10}$ is needed. 
But what does this value mean in comparison to the rate of Eu production by other sources? In Fig. \ref{fig:missEu} we show the  rate of Eu production (normalized to the galaxies SFRs) by MNS and MRD-SNe as compared to the "missing Eu" that explains the observed trends in LG dSphs.

As it stands, it is clear that \emph{to explain the high [Eu/Fe] values observed in these galaxies we need to boost the Eu production rate by a factor of $\simeq 4$} relative to the combined contribution of MRD-SNe and MNS that we use to explain the data in the MW.
In fact, MNSs start to significantly pollute the ISM only at late times (and, therefore, large metallicities) due to their typical delay-times (see \ref{ss:rprocess}), with a contribution dependent on the different SFH shape in different galaxies (see the different line colors in Fig. \ref{fig:missEu}). On the contrary, prompt MRD-SNe naturally track the SFR (which implies no differences in the normalized production rates among galaxies), but they are not sufficient to explain the [Eu/Fe] ratios observed in the three dwarf galaxies.

\section{The origin of the missing Eu}
\label{s:testing_scenarios}

In this Section, we investigate in more detail the "missing Eu" problem. In particular, we suggest a novel Eu enrichment scenario that explains the abundance ratios observed in LG dwarfs in light of the two classes of candidates for Eu production, namely, delayed sources (MNS) and prompt ones (here represented by MRD-SNe). 
After devising such an alternative scenario, we test it against the abundance trends observed in the MW. In this way, we can probe whether or not we can attribute the different patterns observed in different galaxies not only to their different SFH (as due to the time-delay model), but also to intrinsic differences in the properties of their stellar populations (e.g. fraction of binary systems leading to MNS).

\subsection{A larger MNS contribution at low metallicity?}
\label{ss:MNS_add}

To reproduce the large abundance of Eu observed in LG dSphs, here we test the hypothesis of an increased contribution to Eu production by MNS at low metallicities, possibly due to an increased rate of formation of MNS progenitor systems (however, an increased yield per MNS event can also contribute to the Eu boost).
The hypothesis of an increase in the number of MNS progenitor has already been made in other works about the chemical evolution of Eu and r-process in general (e.g. \citealt{Simonetti19,Cavallo22}) and stems from results of binary population synthesis models (e.g. \citealt{Bogomazov07,Mennekens14}), showing that metallicity plays a crucial role in the formation of compact-object binaries (\citealt{Giacobbo18}). 
However, previous assumptions on the dependence of the MNS rate on metallicity did not rely on a quantitative analysis of the available data samples.

Here, we try to fill this gap by running a grid of models for Sgr and For (for Scl, its limited metallicity distribution does not allow to perform a meaningful likelihhod analysis, see later in the Section)  
increasing the fraction of systems in the right mass interval giving rise to MNS events, i.e. $\alpha_{MNS}$, up to a given metallicity threshold, [Fe/H]$_{\rm thresh}$.
In particular, we vary the $\alpha_{MNS}$ parameter introduced in Eq. \eqref{eq:MNS_rate} in this way:
\begin{equation}
    \alpha_{MNS} = \left\{
    \begin{array}{ll}
 2\cdot 10^{-3} \times \alpha_{\rm incr} \hspace{0.52cm} {\rm if} \, \, {\rm [Fe/H]}\leq {\rm [Fe/H]_{thresh}}\\[0.1cm]
 2\cdot 10^{-3} \hspace{1.5cm} {\rm if} \, \, {\rm [Fe/H]} > {\rm [Fe/H]_{thresh}}\\
\end{array}, \right.
\label{eq:alpha_incr}
\end{equation}
where $\alpha_{\rm incr}$ is a multiplicative factor applied to the $\alpha_{MNS}=2\cdot 10^{-3}$ parameter introduced in Tab. \ref{tab:params_rproc}. Concerning the gridding scheme, we adopt an array with
spacing of 0.05 dex for [Fe/H]$_{\rm thresh}$ and $2$ for $\alpha_{incr}$, respectively.
It is worth noting that the step function in metallicity adopted in Eq. \eqref{eq:alpha_incr} may be seen as a simplistic approximation, as a more gradual decreasing trend is expected for the fraction of these systems (see, e.g. \citealt{Giacobbo18}). However, we stick to the above approximation as our goal is to give a first-order quantitative estimate for the decrease of the $\alpha_{MNS}$ parameter. Considering more elaborated functional forms is beyond the scope of this paper, as it would require more/better data.

\begin{figure}
    \centering
    \includegraphics[width=1.\columnwidth]{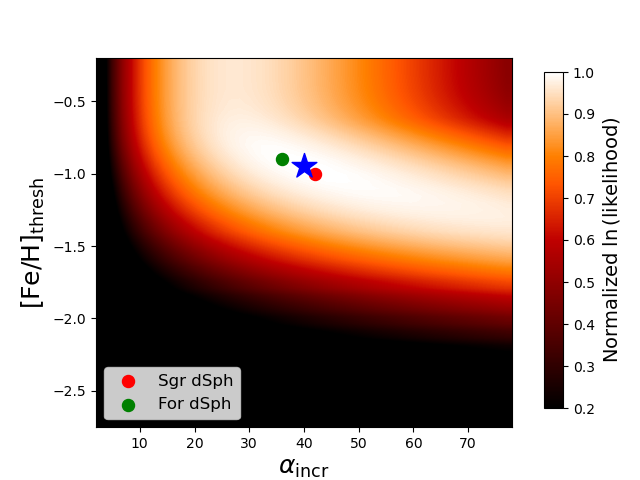}
    \caption{Total weighted likelihood for the sample dwarf galaxies as function of the increase factor in the $\alpha_{MNS}$ parameter $\alpha_{\rm incr}$ and metallicity threshold [Fe/H]$_{\rm thresh}$. Symbols legend is as in Fig. \ref{fig:missEu_likel}.}
    \label{fig:max_likelihood}
\end{figure}

We proceed as done in Section \ref{s:missing_Eu} and compute the likelihood for the grid of models relative to the observed abundance trends (see Appendix \ref{a:Max_Like}).
The results of the likelihood computation are shown in Fig. \ref{fig:max_likelihood}.
It can be appreciated that for both For and Sgr dSph the $\alpha_{MNS}$ parameter has to be multiplied by a large multiplication factor ($\alpha_{incr}> 30$) up to a metallicity of [Fe/H]$\simeq -1$ dex.
Here, the lower metallicity threshold (it was [Fe/H]$_{\rm thresh}\simeq -0.5$ dex in the case of the "missing Eu" rate estimation; see previous section) can be explained in the light of the delayed nature of MNS. In fact, MNS progenitor systems forming at [Fe/H]$<$[Fe/H]$_{\rm thresh}$ can pollute the ISM at much larger metallicities than the adopted threshold.
Concerning the individual galaxies, For shows an increase in MNS extended up to larger [Fe/H] relative to Sgr ($-$0.9 dex and $-$1 dex, respectively), but with relatively lower $\alpha_{incr}$ values (36 and 42, respectively). 
The lower $\alpha_{incr}$ seems in contradiction with the larger "missing Eu" rate found for For relative to Sgr ($7\cdot10^{-10}$ and $4.5\cdot 10^{-10}$ respectively, see Fig. \ref{fig:missEu_likel}). 
However, such a difference can be explained by the larger contribution to Eu production by MNS in For relative to Sgr for metallicities above [Fe/H]$>-1$ dex (see Fig. \ref{fig:missEu}): in fact, the weight of an increased MNS rate on the Eu production will be larger in For relative to Sgr, therefore allowing slightly lower $\alpha_{incr}$ values.
In any case, it is worth reminding that the (small) difference we found can be also partly due to the limited size of the data samples and/or uneven metallicity sampling (see, e.g., Fig. \ref{fig:std_dwarfs}, central panel).
Therefore, we can conclude that the best values found for the two galaxies are  compatible with each other.

\begin{figure*}
    \centering
    \includegraphics[width=0.95\textwidth]{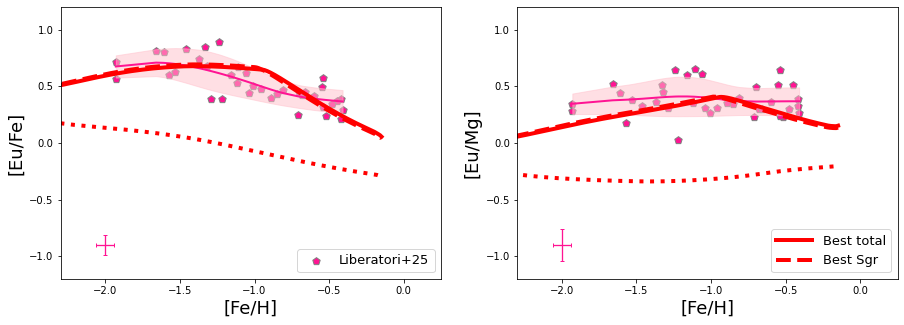}\\
    \includegraphics[width=0.95\textwidth]{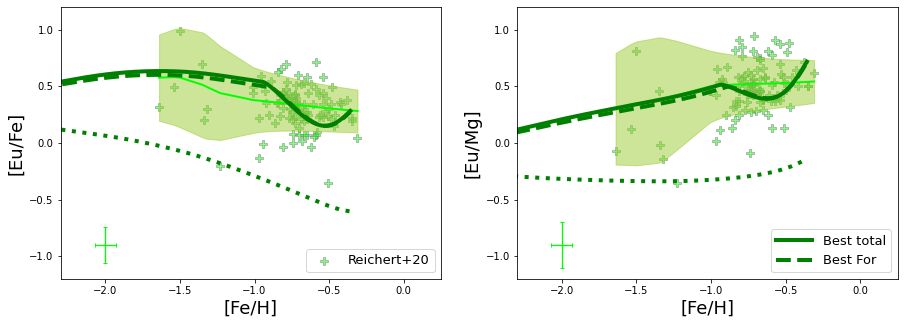}
    \caption{[Eu/Fe] vs. [Fe/H] (left panels) and [Eu/Mg] vs. [Fe/H] (right panels) for Sagittarius (top panels) and Fornax (bottom panels) for models with increased MNS production at low metallicity. 
    Solid lines represent evolutionary tracks for models with Eu enrichment setup with maximum total weighted likelihood, while dashed lines models with maximum likelihood for the individual galaxy. For comparison, the models with standard Eu enrichment setup are also shown (dotted lines). 
    Data legend is as in Fig. \ref{fig:std_dwarfs}.}
    \label{fig:new_dwarfs}
\end{figure*}

The [Eu/Fe] and [Eu/Mg] vs. [Fe/H] patterns obtained by the best models (both total weighted, solid lines, and for individual galaxies, dashed lines) as found by the maximum likelihood analysis are shown in Fig. \ref{fig:new_dwarfs}. These are compared with the "standard" scenario discussed in Section \ref{s:std_scenario} (dotted lines).
The large increase in the fraction of systems giving rise to MNS events allows to reconcile the model predictions for both For and Sgr with the observed trends, which are instead severely underestimated in the case of the "standard" scenario. 
Reassuringly, we note that the observed trends are well recovered both by the best models for the individual galaxies and by the model with parameters from the maximum total weighted likelihood. 
This suggests that, rather than focusing on details such as, e.g., the exact value of the $\alpha_{incr}$ parameter, the reader should fix their attention on the key result of our analysis, that is, \emph{the need for a very large boost in Eu production by MNS at low metallicities.}

At variance with the approach adopted in Sect.~\ref{s:missing_Eu}, we do not perform the same analysis done for Sgr and For for Scl dSph, as the limited metallicity distribution of  Scl (ending at [Fe/H]$\sim-1$ dex) does not allow to obtain a meaningful result from the likelihood computation (see Fig. \ref{fig:missEu_likel}).
Nonetheless, it is fundamental to apply the best model derived for Sgr and For to Scl, to check the consistency of the proposed enrichment scenario for Eu in LG dwarfs.

\begin{figure*}
    \centering
    \includegraphics[width=0.95\textwidth]{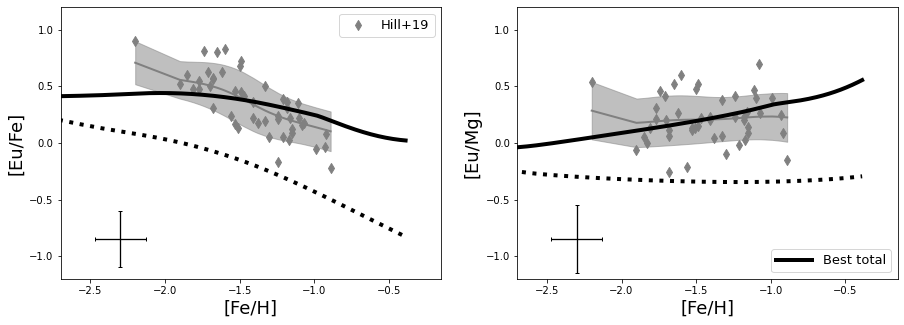}
    \caption{[Eu/Fe] vs. [Fe/H] (left panel) and [Eu/Mg] vs. [Fe/H] (right panel) for Sculptor for models with increased MNS production at low metallicity. 
    Solid lines represent evolutionary tracks for the model adopting the Eu enrichment setup with maximum total weighted likelihood for Sagittarius and Fornax. 
    For comparison, the model with standard Eu enrichment setup is also shown (dotted lines). 
    Data legend is as in Fig. \ref{fig:std_dwarfs}.}
    \label{fig:new_Scu}
\end{figure*}

The [Eu/Fe] and [Eu/Mg] vs. [Fe/H] patterns obtained for Scl dSph with the Eu enrichment setup as found by the maximum likelihood analysis on Sgr and For is shown in Fig. \ref{fig:new_Scu}.
Also in the case of Scl, the new scenario (solid lines) agrees within the 1$\sigma$ confidence interval of the observed trends for [Fe/H]$\gtrsim -2$ dex. In contrast, in the "standard" scenario (dotted lines), the general trend is severely underestimated for both [Eu/Fe] and [Eu/Mg].  
It is also worth reminding that here we focus on the trends for [Fe/H]$\gtrsim -2$ dex (see Section \ref{s:model}), as at lower metallicities stochastic effects heavily affect the observed abundance patterns (e.g. \citealt{Cescutti15,Schoenrich19}). Moreover, in Scl the [Fe/H]$\lesssim -2$ dex range is probed only by a single star \citep[][see also Appendix \ref{a:other_datasets} for other datasets]{Hill19} and therefore slight discrepancies in this metal regime should not worry us much.

\subsection{Prompt sources at low metallicity?}
\label{ss:MRD_Coll_add}

The other possibility we consider to explain the abundance trends in LG dwarf galaxies is a larger contribution from prompt sources, here represented by MRD-SNe (e.g. \citealt{Winteler12,Nishimura15,Nishimura17}). Nonetheless, in this Section we also check possible yields and rates from collapsars, as they represent a common framework for prompt r-process sources in chemical evolution studies (e.g. \citealt{Siegel19,Prantzos20}).

For their prompt nature, such events closely follow the SFR in galaxies: for this reason we do not need to repeat the analysis performed for MNS and consider the "missing Eu" production rate and metallicity threshold found in Section \ref{s:missing_Eu}.
The "missing Eu" rate shown in Fig. \ref{fig:missEu} translates in an increase in the Eu production by prompt events of a factor of $\simeq$ 4.4. Considering the results for individual galaxies, we find a range between $\simeq$3.1 (for Scl dSph) and $\simeq$6.1 (for For dSph).\\

The boosted Eu production could be due either to an augmented event rate or to an increased Eu yield. 
In particular, if we assume a fixed $\alpha_{MRD}=$ 0.2 (see Tab. \ref{tab:params_rproc}), an increase in the Eu production by a factor of 4.4 implies an (average) Eu yield by MRD-SNe of $\simeq 2\times 10^{-6}$ M$_\odot$. Similar values are obtained in MRD-SNe simulations assuming larger magnetic field strength (e.g. model L0.6 of \citealt{Nishimura17}), therefore leaving the door open for larger yield values.
On the other hand, if we assume a total r-process production per collapsar $>$0.1 M$_\odot$ (\citealt{Siegel19}), resulting in an Eu yield of $\gtrsim 10^{-5}$ M$_\odot$ (assuming a solar scaled r-process pattern, \citealt{Arnould07}), this would imply that $\lesssim$5$\%$ of very massive stars ($m>20$ M$_\odot$) are progenitors of collapsar events. Again, such values are theoretically possible according to the ratio of SN Ib/c and long GRB rate\footnote{The collapsar model (\citealt{MacFadyen99,Woosley06}) is one of the favorite scenarios for long GRB events, originating from the explosion of a single, very massive ($m>>10$ M$_\odot$) rotating star that collapses to form a rapidly rotating black hole.} ($\lesssim 1\%$, \citealt{Ghirlanda22} and references therein) and the observed fraction of SN Ib/c which 
correspond to very massive stars and not stripped binaries ($\simeq 10 \%$, \citealt{Karamehmetoglu23}).

It is clear that all the solutions to the problem of Eu underproduction in LG dwarf galaxies invoking an increased contribution to Eu enrichment from prompt sources are highly degenerate. 
The safest fix at our stage rests, perhaps, on the possibility of an increased prompt event rate so as to match the observations in LG dwarfs.

\subsection{Applying the new scenarios to the MW}
\label{ss:new_scenario_MW}

In the previous paragraphs, we have advocated a boost in Eu production from either prompt or delayed stellar sources below a given metallicity threshold to match the theoretical predictions of our chemical evolution models for LG dSphs to the observations. In the remainder of this section, we will see whether our new Eu enrichment scenario can be applied to the MW, without leading to inconsistencies with the data.

\begin{figure*}
    \centering
    \includegraphics[width=0.95\textwidth]{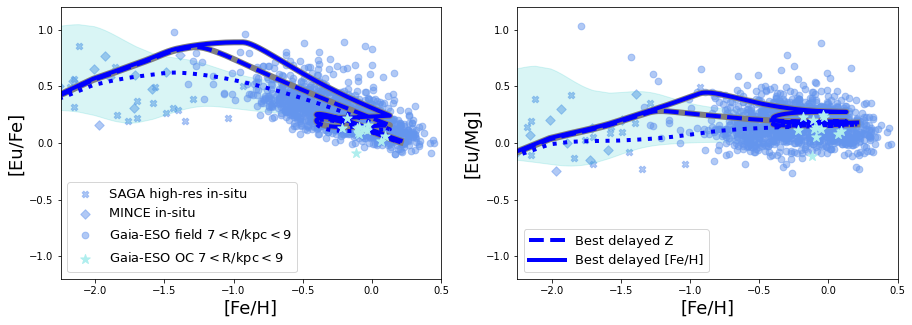}
    \caption{[Eu/Fe] vs. [Fe/H] (left panel) and [Eu/Mg] vs. [Fe/H] (right panel) for the MW for models with increased MNS production at low metallicity.
    Solid lines represent evolutionary tracks for models with Eu enrichment setup as the best model for Local Group dwarfs, 
    while dashed lines models with metallicity threshold equal to a [Fe/H] for which the MW galaxy has a metallicity $Z$ corresponding to the [Fe/H]$_{thresh}$ threshold as computed for Local Group dwarfs. 
    For comparison, the model with standard Eu enrichment setup is also shown (dotted lines). 
    Data legend is as in Fig. \ref{fig:std_MW}.}
    \label{fig:new_MW}
\end{figure*}

The [Eu/Fe] and [Eu/Mg] vs. [Fe/H] abundance patterns for the MW model with an increased Eu production by MNS systems formed at low metallicity (with parameters from the best model for dwarf galaxies, i.e. $\alpha_{\rm incr}=40$, [Fe/H]$_{\rm thresh}=-0.95$ dex) are shown with the solid lines in Fig. \ref{fig:new_MW}. 
As expected, for both [Eu/Fe] and [Eu/Mg] the model shows a peak approximately at [Fe/H]$\simeq-0.9/-0.8$ dex $\gtrsim$ [Fe/H]$_{\rm thresh}$.
However, the peak [Eu/X] ratios are well above the confidence interval of the data trend. This is especially the case for [Eu/Fe] (Fig. \ref{fig:new_MW} left panel), where the model predictions are up to $\sim$0.3-0.4 dex above the upper end of the confidence interval of the observed trend.

However, by looking at the [Fe/H]$_{\rm thresh}$ values obtained from the maximum likelihood analysis on the individual Sgr and For dSphs (see Fig. \ref{fig:max_likelihood}, red and green points), [Fe/H]$=-1$ dex and [Fe/H]$=-0.9$ dex, respectively, we note 
that they correspond to similar global metallicities, namely Z$\simeq$ 2.2$\times 10^{-3}$ for Sgr and Z$\simeq$ 1.8$\times 10^{-3}$ for For. In turn, these translate into solar-scaled metallicities between $\simeq$ 0.105 and 0.13  Z$_\odot$ (assuming Z$_\odot$ from \citealt{Grevese98}) or $\simeq$ 0.13 and 0.16 Z$_\odot$ (assuming Z$_\odot$ from \citealt{Asplund09}).

On this regard, it is worth saying that the differences between the SFHs of dSphs and more star forming systems lead to different chemical enrichment patterns. 
This is due to the time-delay model (e.g. \citealt{Matteucci12,Matteucci21}), for which the evolution of [X/Fe] vs. [Fe/H] patterns are different depending on the galaxy star formation. As $\alpha$-elements, such as oxygen, show enhanced/depleted [$\alpha$/Fe] at a given [Fe/H] in systems with larger/lower star formation rate and being oxygen the metal dominating the global metallicity $Z$ budget, variations in the galaxy star formation rate influence the relation between [Fe/H] and $Z$.
Therefore, the MW galaxy should exhibit the $Z$ metallicities corresponding to the [Fe/H]$_{thresh}$ of dwarf systems at lower [Fe/H].
Indeed, we find that for our MW model, [Fe/H]$\simeq -$1.33 dex at Z$\simeq$ 2$\times 10^{-3}$.
Therefore, we run an additional model for the MW where we assume the same $\alpha_{\rm incr}$ factor as found from the best model in Section \ref{s:testing_scenarios}, but assuming a lower [Fe/H]$_{\rm thresh}$, namely $-$1.33 dex, corresponding to a $Z\simeq2\times10^{-3}$ in the MW.
The results for this additional model are displayed with dashed lines in Fig. \ref{fig:new_MW}. Now, the model predictions align much better with the observed trends, lying within the $1\sigma$ confidence interval in [Eu/Mg] and just slightly above it ($\lesssim$0.1 dex) in [Eu/Fe]. In this context, it is also important to emphasize that when discussing metallicity, we should carefully distinguish between using global $Z$ and the stellar [Fe/H].

The important uncertainties in the modelling of r-process (yields, nucleosynthetic sources fraction within the IMF, see e.g. \citealt{Molero21} and references therein) even for our Galaxy imply that such marginal disagreement in the trends can be recovered with few minor adjustments in some of the free r-process parameters (e.g. $\alpha_{MRD}$, MRD yields) without compromising the reproduction of observables (e.g. the present-day MNS rate).
Moreover, the small discrepancy shown by this model may be also explained by uncertainties in the derivation of Eu stellar abundances, with ${\rm<3D>}$ and NLTE corrections that increase the median [Eu/Fe] trend in the MW of around $\sim$0.1 dex at [Fe/H]$\sim$-1.5/-1 dex (see, e.g. Fig. 8 by \citealt{Guo24}).
The same line of reasoning cannot be applied to the model shown with the solid lines in Fig. \ref{fig:new_MW}, because of the much more severe offset (up to $\sim 0.3-0.4$ dex) with the observations. \\

\begin{figure*}
    \centering
    \includegraphics[width=0.95\textwidth]{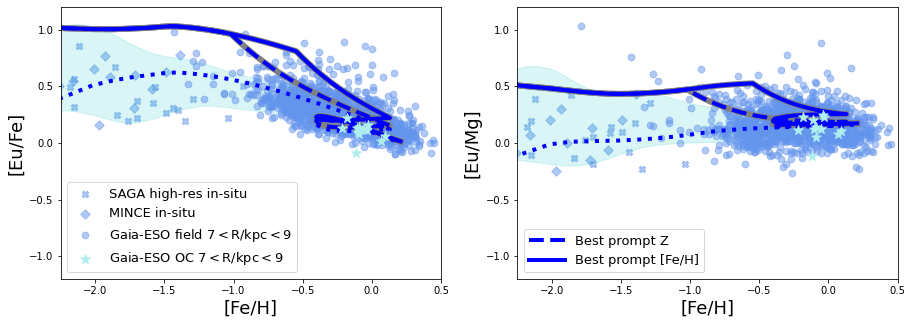}
    \caption{[Eu/Fe] vs. [Fe/H] (left panel) and [Eu/Mg] vs. [Fe/H] (right panel) for the MW models with increased prompt sources (MRD-SNe and/or collapsars) production at low metallicity.
    Solid lines represent evolutionary tracks for models with Eu enrichment setup as the best model for Local Group dwarfs, 
    while dashed lines models with metallicity threshold equal to a [Fe/H] for which the MW galaxy has a metallicity $Z$ corresponding to the [Fe/H]$_{thresh}$ threshold as computed for Local Group dwarfs.
    For comparison, the model with standard Eu enrichment setup is also shown (dotted lines). 
    Data legend is as in Fig. \ref{fig:std_MW} and \ref{fig:new_MW}.}
    \label{fig:new_MW_ADD}
\end{figure*}

We repeat the kind of analysis outlined above in the case of a putative increased production of Eu from prompt sources (MRD-SNe and/or collapsars) at low metallicity.
In particular, the solid lines in Fig. \ref{fig:new_MW_ADD} refer to a MW model with parameters from the best-fitting model for dwarf galaxies ("missing Eu" rate$=5\cdot 10^{-10}$, [Fe/H]$_{\rm thresh}=-0.55$ dex, see Section \ref{s:missing_Eu}).
In addition, we show the results of a model with increased prompt sources production, but with a lower [Fe/H]$_{thresh}$, corresponding to the global metallicity $Z$ as found for the [Fe/H] threshold in Sgr and For. 
In particular, we apply a [Fe/H]$_{\rm thresh}\simeq-1.02$ dex, 
corresponding in our MW model to $Z=3.5\times 10^{-3}\simeq0.21$ Z$_\odot$ (assuming Z$_\odot$ from \citealt{Grevese98}), namely in between the values found for Sgr ($Z\simeq4.7\times 10^{-3}$) and For $Z\simeq2.7\times 10^{-3}$ individually.

Looking at the model with metallicity threshold calibrated on [Fe/H] it is clear that the boost of Eu production from prompt sources leads to a general overestimation of the observed Eu content in the Galaxy even beyond the adopted threshold of [Fe/H]= $-$0.55 dex.
For [Eu/Fe], in particular (see Fig. \ref{fig:new_MW_ADD}, left panel), the model predicts a low-metallicity plateau with extreme values ([Eu/Fe]$\simeq$1 dex), which disagrees with the average trend spotted in the data.
The situation is slightly better in the [Eu/Mg] vs [Fe/H] plane (Fig. \ref{fig:new_MW_ADD}, right panel), where the trend predicted by the model remains within the 1$\sigma$ confidence interval of the data up to [Fe/H]$\simeq -$1.5 dex. However, also in this case the high plateau ([Eu/Mg]$\simeq$0.5 dex) prevents the agreement with the bulk of the data at larger metallicities, confirming the overproduction of Eu in the MW by this model.

When we apply the threshold on the global metallicity, $Z$, rather than on [Fe/H], the agreement between model predictions and data does not improve substantially. In fact, despite the shorter plateau predicted for [Eu/Fe] and [Eu/Mg], the model tracks still severely overestimate the observed trend up to [Fe/H]$\gtrsim-1$ dex.
Therefore, the results displayed in Fig. \ref{fig:new_MW_ADD} indicate the inability of the scenario favoring Eu production from prompt events to explain the abundance patterns in the MW, excluding it as a possible way to reconcile the trends in the MW and LG dwarf galaxies.

\section{Discussion and conclusions}
\label{s:discussion_conclusion}

In this study, we investigated the chemical enrichment of the r-process element Europium in galaxies of the Local Group by means of detailed models of chemical evolution, considering both a prompt (among the hypotheses, magneto-rotationally driven supernovae, hereafter MRD-SNe, e.g. \citealt{Winteler12}, or collapsars, e.g., \citealt{Siegel19}) and a delayed (merging of compact objects / neutron stars, hereafter MNS, e.g. \citealt{Argast04}) source of enrichment.

In particular, we focused on the simultaneous modelling of Eu abundance patterns observed in the Milky Way and in the three most massive dwarf Spheroidal galaxies of the LG, namely Sagittarius, Fornax and Sculptor.
This effort is necessary, as Eu has started to be used as a possible chemical tag for accreted structures in the field of Galactic Archaeology (see, e.g. \citealt{Monty24}) but we are currently lacking a detailed theoretical understanding of the evolution of this element in different galactic systems.
In fact, apart from the very metal-poor regime ([Fe/H]$\lesssim -2$ dex), most theoretical work only focused on Eu evolution in the MW galaxy.
To answer to this need, we started from a well-tested model that reproduces all the available observables in the MW disk (\citealt{Molero2023MNRAS.523.2974M}) and expand the same setup for Eu enrichment (nucleosynthesis sources, yields and delay in the pollution for MNS) to models tailored to reproduce the star formation histories, the lighter elements abundance ratios (e.g. [$\alpha$/Fe]) and the metallicity distribution functions of Sgr, For and Scl.  \\

Following this approach, the main conclusions of this work can be summarised as follows:
\begin{itemize}
    
    \item the adopted "standard" setup for Eu enrichment, which well reproduces both the measured abundance trends and the event rates in the MW (see \citealt{Molero2023MNRAS.523.2974M} and Figs. \ref{fig:rates_MW}, \ref{fig:std_MW}), underestimates significantly the [Eu/Fe] abundance ratio in the three most massive LG dSphs (Sgr, For, Scl).  In particular, the model predictions underestimate the observed [Eu/Fe] trend by $\sim$0.5 dex for all three galaxies (see Fig. \ref{fig:std_dwarfs});

    \item we quantify the amount of "missing Eu" and its extent in metallicity in LG dwarf galaxies. 
    To reproduce the observed trends in LG dSphs we need an additional Eu production rate (normalised to the SFR) of $\simeq 5 \cdot 10^{-10}$ up to metallicities [Fe/H]$\simeq-$0.5/$-$0.6 dex, with similar results for the different galaxies (see Fig. \ref{fig:missEu_likel}). 
    By comparing the result with the production rates from MNS and MRD-SNe, we find that a boost of a factor $\simeq 4$ in the Eu production rate is needed to reconcile the model predictions with the observations;

    \item searching for the origin of this "missing Eu", we explored two possible scenarios:
    i) a larger contribution by MNS systems (or delayed scenario) or ii) an increment in the production by MRD-SNe or collapsars (prompt scenario), both at low-metallicity.
    
    For the delayed scenario, we find that we need to increase the produced Eu from MNS by a factor of $\simeq$40 from the systems forming at [Fe/H]$\lesssim-1$ dex (see Fig. \ref{fig:max_likelihood}).
    For the prompt scenario, the increase in Eu production we propose ($\simeq 4.5$) is compatible with theoretical values found by both MRD-SNe (e.g. \citealt{Nishimura17}) and collapsars models (\citealt{Siegel19,Ghirlanda22,Karamehmetoglu23}) and needs to extend up to [Fe/H]$\lesssim-0.6$ dex;

    \item both the delayed and the prompt scenarios that explain the LG dwarf galaxy data are applied to the chemical evolution model for the MW, to check whether or not these scenarios are compatible with the chemical trends found in the Solar Neighbourhood.
    The resulting [Eu/Fe] and [Eu/Mg] vs. [Fe/H] abundance patterns (see solid lines in Figs. \ref{fig:new_MW} and \ref{fig:new_MW_ADD}) are generally above the observed trends, with the MW model adopting the prompt scenario overestimating the observed trend up to $\sim$0.4 dex and the delayed scenario by $\sim$0.3 dex;

    \item nonetheless, if we consider the global metallicity $Z$ instead of [Fe/H] as a threshold for the delayed scenario (see dashed lines in Fig. \ref{fig:new_MW})\footnote{ in practice, we consider a [Fe/H] threshold for which the MW galaxy has a metallicity $Z$ corresponding to the [Fe/H] threshold as computed for LG dwarfs.}, the model predictions are in fairly good agreement with the observations, with discrepancies ($\lesssim 0.1$ dex) that may be ascribable to limitations in the modelling of r-process elements chemical enrichment (e.g. \citealt{Cote19,Molero21,Molero2023MNRAS.523.2974M}) or derivation of Eu abundances from spectral lines (e.g. \citealt{Mashonkina17,Guo24}).
    
    In turn, this suggests the delayed scenario as a possible solution to theoretically reconcile the trends observed in the MW and in LG dwarfs without advocating intrinsic differences in the stellar populations of the galaxies.
    Moreover, the present analysis emphasize that in discussing metallicity, we should carefully distinguish between using global Z and the stellar [Fe/H] when discussing the properties of stellar populations, such as MNS progenitor systems. 
    
\end{itemize}

Summarizing, our results proved that an increased production of Eu by MNS systems forming at low metallicity (Z$\lesssim0.1$ Z$_\odot$) may be seen as a viable scenario to explain the Eu enrichment of different galaxies in the LG.  
However, due to the severe degeneracies still affecting the models, we cannot reach a conclusive assessment of Eu evolution in LG galaxies. On one side, despite the statistically meaningful approach adopted here, the limited sizes of the samples of stars with Eu abundance determinations in LG dwarfs prevent us to capture small variations in the parameters characterizing the proposed enrichment scenarios. On the other hand, the modelling of the chemical evolution of r-process elements in general presents multiple challenges, with extremely limited knowledge and ambiguities persisting even in the most basic ingredients for chemical evolution models. As an example, stellar yields from candidate Eu polluters are extremely poorly sampled in terms of progenitors masses and metallicity, so that chemical evolution models still have to rely on single point grids (see also \citealt{Molero25} for a discussion on this point).

Nonetheless, this work highlights the importance of considering galactic ecosystems different from the MW in the modelling of neutron-capture element evolution and in particular Eu, which is the main tracer of r-process production. 
In fact, by looking at galaxies with different SFHs we can mitigate the degeneracies in the r-process modelling and shed light on the nature of the different nucleosynthetic sources. \\

Further targeting of neutron-capture elements in LG galaxies (also including, e.g. the more massive and interacting Magellanic Clouds) is certainly needed to fill the gaps in our understanding, building samples that are statistically significant and complete over the metallicity range spanned by LG galaxies. In this regards, upcoming facilities, such as 4MOST (\citealt{4MOSTpaper}), will reserve significant amount of time on the galaxies object of this study, allowing to increase the number of observed stars with abundances of neutron-capture elements by orders of magnitude (\citealt{Skula4MOST}).

\begin{acknowledgements}
The authors thank the referee for the careful reading of the manuscript and the comments provided. 
MP thanks F. Matteucci for the insightful discussions and suggestions on the paper content. 
MP, DR, and AM acknowledge financial support from the project "LEGO – Reconstructing the building blocks of the Galaxy by chemical tagging" granted by the Italian MUR through contract PRIN2022LLP8TK\_001. 
MM thanks the support from the Deutsche Forschungsgemeinschaft (DFG, German Research Foundation) – Project-ID 279384907 – SFB 1245, the State of Hessen within the Research Cluster ELEMENTS (Project ID 500/10.006).
    
\end{acknowledgements}

%
   \bibliographystyle{aa} 
   \bibliography{paper_FINAL} 
%

\appendix

\section{Maximum likelihood calculations}
\label{a:Max_Like}

In the main text we show the results of the likelihood computation performed on the grid of parameters $\Theta=\{{\rm "Missing\,\, Eu"\,\, rate}\, , \, {\rm [Fe/H]}_{\rm thresh} \}$ (in the "missing Eu" case, Section \ref{s:missing_Eu}) and $\Theta=\{\alpha_{\rm incr} \, , \, {\rm [Fe/H]}_{\rm thresh} \}$ (in the MNS increased production case, Section \ref{ss:MNS_add}).
In doing this, we adopt a similar procedure to what illustrated in \citet{Spitoni20,Spitoni21}. 
In the following, we show the details on how we perform these computations.\\

For each individual galaxy, the set of observables is $x_{gal} = \{{\rm [Fe/H] \, , \, [Eu/Fe] \, , \, [Eu/Mg]}\}$, while the set of model parameters is $\Theta = \{var_1 \, ,  \, var_2 \}$ (where $var_1$ and $var_2$ are as in the cases illustrated above). 
The likelihood, $\mathcal{L}$ , assuming that the uncertainties on the observables are normally distributed, can be therefore written as:
\begin{equation}
    \ln \mathcal{L}_{gal} \, = \, - \sum_{n=1}^{N_{gal}} \ln \Bigg( (2\pi)^{d/2} \prod_{j=1}^{d} \sigma_{n,j} \Bigg) - \frac{1}{2} \sum_{n=1}^{N_{gal}} \sum_{j=1}^d \Bigg( \frac{x_{n,j} - \mu_{n,j}}{\sigma_{n,j}} \Bigg)^2,
    \label{eq:likelihood}
\end{equation}
where $N$ is the number of stars for the individual galaxy. The quantities $x_{n,j}$ and $\sigma_{n, j}$ are respectively the measured value of $j$-th observable and its uncertainty for $n$-th star in the individual galaxy, whereas $\mu_{n, j}$ is the model value for to the $j$-th observable associated to the $n$-th star.

To associate to each star the corresponding predicted value by the chemical evolution model we compute the closest value on the curve given a data point $x_{n, j}$ as in \citet{Spitoni20}, namely defining the following $\lq$data-model distance' function $D$:
\begin{equation}
    D_{n,j} \, = \, \sqrt{ \sum_{j=1}^d \Bigg( \frac{x_{n,j} - \mu_{n,j,i}}{\sigma_{n,j}} \Bigg)^2},
    \label{eq:distance_datamodel}
\end{equation}
where $i$ runs over a set of discrete values on the curve. Hence, the closest point on the curve will be $\mu_{n, j}$ = $\mu_{n, j,i0}$, which fulfils the following condition:\\
\begin{equation}
    S_n  \, \equiv \,  \min_i\{ D_{n,j} \} \, = \,  \sqrt{ \sum_{j=1}^d \Bigg( \frac{x_{n,j} - \mu_{n,j,i0}}{\sigma_{n,j}} \Bigg)^2}. 
    \label{eq:min_distance_datamodel}
\end{equation}\\

After the computation of the likelihood for the individual galaxies, we then calculate a "total weighted" likelihood over the set of galaxies.
To do that, we weight the log likelihood for each individual galaxy according to the number stars observed within it, as shown in the following equation:
\begin{equation}
    {\rm Weighted}\, (\, \ln \mathcal{L}_{tot} \,) = \sum _{gal} \ln \mathcal{L}_{gal} \times N_{gal}.
    \label{eq:global_likelihood}
\end{equation}
In this way, we avoid to underweight or overweight the likelihood for the individual galaxies accordingly to the different sizes of the samples (e.g $\sim$ 40 stars for Sgr, $\sim$ 110 for For).

\section{Testing on additional datasets for local dwarf galaxies}
\label{a:other_datasets}

As we mention in Section \ref{s:testing_scenarios}, we compare the best models emerged from our likelihood analysis (see Appendix \ref{a:Max_Like}) to other available datasets for the LG dwarf galaxies analysed in this work.

\begin{figure}
    \centering
    \includegraphics[width=0.95\columnwidth]{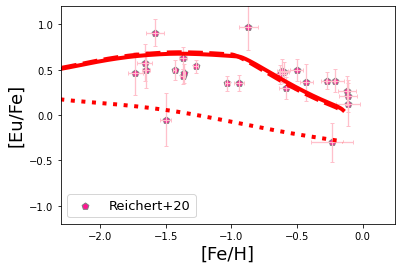}\\
    \includegraphics[width=0.95\columnwidth]{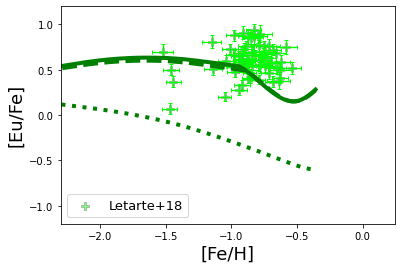}\\
    \includegraphics[width=0.95\columnwidth]{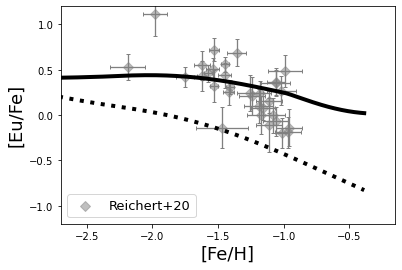}
    \caption{[Eu/Fe] vs. [Fe/H] for Sagittarius (top panel), Fornax (central panel) and Sculptor (bottom panel) for models with increased MNS production at low metallicity and different datasets to the ones described in Section \ref{ss:dataLG}.
    Data are from \citealt{Reichert20} (Sagittarius and Sculptor) and \citet{Letarte10,Letarte18} (Fornax).
    Model legend is the same as in Figs. \ref{fig:new_dwarfs} and \ref{fig:new_Scu}.}
    \label{fig:other_dataset}
\end{figure}

In particular, here we consider the datasets from \citet{Reichert20} for Sgr and Scl and the one by \citet{Letarte10,Letarte18} for For dSph. 
These additional datasets are not considered in our main analysis as due to the disadvantages they present relative to the adopted samples.
These are, for example, the smaller samples with available Eu abundances (Sgr, For, Scl), the less careful stellar selection (Sgr), 
the lower spectral resolution in the observations (Sgr, Scl). 
Moreover, the non-negligible overlap in the catalogues of For and Scl ($\sim 70$ and $\sim 35$ stars, respectively) does not allow to increase significantly the number of stars observed in each galaxy, and therefore their addition would not be beneficial for the study.\\

The resulting best models with increased Eu production from MNS (see Section \ref{ss:MNS_add}), together with the ones adopting the standard scenario for Eu enrichment, are compared with these other datasets in Fig. \ref{fig:other_dataset}.

This Figure confirms the findings in the main text, with the standard scenario unable to fulfil the Eu production needed in the different LG dwarfs.
Indeed, models with an additional r-process production component at low metallicity (in this case, from MNS) are necessary to fill the gap with [Eu/Fe] observations.

\end{document}